\documentclass{aa}
\usepackage{natbib}
\usepackage{graphics}
\usepackage[mathcal]{eucal}
\usepackage{amssymb} 

\newcommand{\fig}[3]{
      \begin{figure}[tbp]
        \resizebox{\hsize}{!}{\includegraphics  {#1}}
        \caption{#2}
        \label{#3}
        \end{figure} }
\newcommand{\eqn} [1] {
\begin{equation}#1
\end{equation}}
\newcommand{\eqna} [1] {
\begin{eqnarray}#1
\end{eqnarray}}
\newlength{\lenA} %
\begin{document}
\title{Solar-like oscillations in $\delta$~Scuti stars}
\author{Samadi R. \inst{1,2}\and Goupil M.-J.\inst{2} 
                            \and Houdek G.   \inst{3}}

\institute{
Astronomy Unit, Queen Mary, University of London, London E14NS, UK \and
Observatoire de Paris, LESIA, CNRS FRE 2461, 92195 Meudon, France \and
Institute of Astronomy, University of Cambridge, Cambridge CB3 0HA, UK 
}
\offprints{R. Samadi}
\mail{Reza.Samadi@obspm.fr}
\date{Received / Accepted}

\abstract{
Model computations of $\delta$~Scuti stars, located in the vicinity of the red 
edge of the classical instability strip, 
suggest amplitudes of solar-like oscillations larger than in cooler 
models located outside the instability strip. 
Solar-like amplitudes in our $\delta$~Scuti models are found to be large 
enough to be detectable with ground-based instruments provided 
they can be distinguished from the opacity-driven large-amplitude pulsations.
There would be advantages in observing simultaneously opacity-driven and
stochastically excited modes in the same star. We anticipate their possible
detection in the context of the planned asteroseismic space missions,
such as the French mission COROT (COnvection ROtation and planetary Transits). 
We propose known $\delta$~Scuti stars
as potential candidates for the target selection of these upcoming 
space missions.
}
\maketitle

\section{Introduction}

The $\delta$~Scuti stars are in general main sequence stars with masses 
between 1.5\,M$_\odot$ and 2.5\,M$_\odot$. 
They are located  inside the classical instability strip (IS hereafter) 
where the $\kappa$-mechanism drives low-order radial and nonradial modes of 
low degree to measurable amplitudes (opacity-driven unstable modes).  Only a 
small number of opacity-driven modes are observed in $\delta$~Scuti stars
\citep[for a review see e.g.][]{Gautschy96},
but their amplitudes, which are limited by nonlinear processes, are much 
larger than stochastically driven intrinsically stable solar-like p modes.

For main-sequence stars with surface convection zones, located outside the IS, 
model computations suggest all modes to be intrinsically stable but excited 
stochastically by turbulent convection; for models located near the red edge 
of the IS the predicted velocity 
amplitudes become as large as 15 times the solar value \citep{Houdek99}.
Moreover, these computations suggest that models located inside the IS
can pulsate simultaneously with modes excited both by the $\kappa$-mechanism
and by the turbulent velocity field.

{ 
Provided that many modes can be detected, high-frequency p modes are more easily 
identified than low frequency p modes. Hence there are advantages of observing 
simultaneously both types of modes in the same star. As a first step, high-frequency 
p modes can help to determine the fundamental stellar parameters (e.g., luminosity, 
effective temperature) more accurately,
whereas low-frequency modes, which are strongly sensitive to the
properties of the deep layers of the star, can then be used as a diagnostic 
for the inner properties of the model. 
Such developments are outside the scope of the present paper and 
we only outline briefly the underlying idea. 

The nearly regular frequency spacing of solar-like modes of 
high order (i.e., the large frequency separation) depends predominantly on 
the structure of the surface layers and consequently provides further 
constrains on the equilibrium models. 
Their  degree $l$ and azimuthal order $m$ can be identified with the help
of the classical echelle diagram method; this method was successfully tested
by the simulation results of the COROT Seismic Working Group 
(Appourchaux, 2002, personal communication);
this severely constraints the fundamental stellar parameters 
(mass, age, chemical composition) of models for which the frequencies of
computed oscillation modes are similar to the observed high-order modes 
\citep{Berthomieu02}. 
Moreover, solar-type modes also provide information on the star's mean 
rotation rate.

A nearly regular spacing in frequency is also observed for opacity-driven 
low-frequency modes \citep{Breger99}; the large separation of these
low-frequency modes has to be similar between observations and 
theoretical models which satisfy also the properties of the observed
high-frequency solar-type p modes.
However, some of the opacity-driven modes deviate from the mean value of
the large frequency separation; these modes are so-called mixed modes 
which provide details of the stellar core and of the precise evolutionary 
stage of the observed star \citep[][ and references therein]{Unno89}.
This deviation from the mean value of the large frequency separation 
could suggest the presence of mixed modes. The problem is further
complicated by the fact that the rotational splitting frequency components 
are no longer equidistant for these fast rotators, i.e. these frequencies 
could erroneously be identified as frequencies of mixed modes.
However, knowing the mean rotation rate from the high-frequency splittings of
solar-type p modes, the frequency splittings of the low-frequency 
opacity-driven modes can be computed in the manner of 
\citet{Dziembowski92} \citep[see also ][]{Soufi98}.
}

The understanding of the physics responsible for the return to stability of 
opacity-driven modes at the red edge of the IS is still in its infancy.
As the star becomes cooler the extent of the surface convection zone increases,
thereby making the effect of convection-pulsation coupling on mode stability
progressively more important. 
Several authors have tried to model the location of the red edge, e.g., 
\citet{Baker79}, \citet{Bono95} for RR Lyrae stars and e.g., 
\citet{Houdek96} and 
Xiong \& Deng (2001) for $\delta$~Scuti stars. Although the authors assumed 
various models for the time-dependent treatment of convection, they all 
concluded that convection dynamics crucially effect the location of the red 
edge; however, different results were reported as to whether the convective 
heat flux (e.g., \citealp{Bono95}), the momentum flux (e.g., Houdek,\,1996) or
turbulent viscosity (Xiong \& Deng,\,2001) is the crucial agent for 
stabilizing the modes at the red edge.  
In all these investigations, 
the predicted position of the red edge depends crucially on the assumed
convection parameters, such as the mixing-length parameter or whether
acoustic emission is included or neglected in the equilibrium model
\citep{Houdek00}.

%
Although it is possible from Fig.~(13) of \citet{Houdek99} to conclude that 
both types of modes can be excited simultaneously in the same star, amplitudes 
of stochastically excited modes for stars located inside the instability strip
were not explicitly carried out by \citet{Houdek99} and their possible 
detection were not addressed.

The aim of this paper is to demonstrate that models of 
stars, located inside the IS and near the red edge,
can exhibit both opacity driven modes 
and solar-like oscillations with sufficiently large amplitudes to be detectable
with today's ground-based instruments. Consequently 
the planned asteroseismology space missions, such as COROT 
\citep[COnvection ROtation and planetary Transits, ][] {Baglin98} or 
Eddington \citep{Favata00}, will detect these oscillations even more easily.

Sect.~2 describes the equilibrium models, and the linear analysis results
are discussed in Sect.~3., which are obtained from solving the equations of 
linear nonadiabatic oscillations in which convection is treated with the
time-dependent, nonlocal formalism by 
\citet[][hereafter G'MLT]{Gough76,Gough77}. Furthermore, the effect of 
acoustic radiation in the equilibrium model on the stability properties is 
taken into account in the manner of \citet[ and references therein]{Houdek00}. 
In this paper we consider only radial p modes.

Amplitudes of solar-like oscillations result from the balance between damping
and stochastic driving by turbulence. The rate at which the turbulence 
injects energy into the p modes is estimated in the manner of 
\citet[][ Paper~I hereafter]{Samadi00I} and is discussed in Sect.~4.
 
In Sect.~5 we address the possibilities and conditions for detecting
solar-type oscillations in $\delta$~Scuti stars with ground-based 
instruments and propose possible candidates, some of which are listed
in the catalogue by \citet{Rodriguez00}. 
Conclusions are given in Sect.~6.

\section{The stellar models}

Equilibrium envelope models are computed in the manner of \citet{Houdek99}
using G'MLT formulation for convection. Integration starts at an optical
depth of $\tau=10^{-4}$ and ends at a radius fraction 0.2. Radiation is
treated in the Eddington approximation and the atmosphere is assumed to be
grey and plane parallel. 
{ 
In G'MLT formulation two more parameters, $a$ and $b$, are introduced 
which control the spatial coherence of the ensemble of eddies contributing 
to the total heat and momentum fluxes ($a$), and the degree to which the 
turbulent fluxes are coupled to the local stratification ($b$). In this paper
we choose $a^2=900$ and $b^2=2000$ in order to obtain stable modes in the
frequency range in which the damping rates exhibit a local minimum
(e.g., at about 1.1\,mHz for model~C; see Section~3 and Fig.~\ref{fig:cmp_eta}).
}
The mixing-length parameter $\alpha$ has been calibrated to 
a solar model to obtain the helioseismically inferred depth of the convection 
zone of 0.287 of the solar radius \citep{JCD91}. 
\begin{table}
\caption{Stellar parameters for the envelope models~A1, A2, B1, B2 and C; 
$R$ is the stellar 
radius at the photosphere ($T=T_{\rm eff}$), and $\nu_c$ is the acoustic 
cut-off frequency.
}
\label{tab:models}
\begin{center}
\begin{tabular}{cccccc}  
Model &$T_{\rm eff}$&$(b-y)_0$ &$R$        &$\nu_c$&acoustic \cr
      & [K]         &          &[R$_\odot$]& [mHz] &radiation\cr
\hline
A1 , A2 &  $6839$ & 0.235 & 2.40 &1.4 &included\cr
B1 , B2  &  $6839$ & 0.235 & 2.40 &1.4 &neglected \cr
C &  $6650$ & 0.262 & 2.54 &1.3 &neglected \cr
\end{tabular}
\end{center}
\end{table}
\begin{table}
\caption{{Acoustic emissivity coefficient $\Lambda$ and Mach-number dependence 
$\Gamma$ assumed in the acoustic radiation model for the stellar models A1 and 
A2.} }
\label{tab:parameters}
\begin{center}
\begin{tabular}{crl}  
Model  &$\Lambda$\ \ &\ $\Gamma$\cr
\hline
A1 & 100 & 5\cr
A2 & 2000 & 7.5 \cr
\end{tabular}
\end{center}
\end{table}

All models assume solar chemical composition and have mass 
$M=1.68\,$M$_\odot$ and luminosity $L=11.3\,$L$_\odot$, but differ 
in effective temperature $T_{\rm eff}$, and whether or not acoustic 
radiation is included in the equilibrium computations.
Table~\ref{tab:models} lists the fundamental stellar parameters 
of these models. The models A1, A2 , B1 and B2 are hotter than model~C 
and are located inside the IS and close to the red edge. 
{Models A1 and A2 differ from models B1 and B2 by the inclusion of acoustic 
radiation by turbulence in the envelope calculations.
In this model for acoustic radiation in the equilibrium model two more
parameters are introduced \citep{Houdek98}: the emissivity coefficient
$\Lambda$ and the parameter $\Gamma$ which describes the power-law dependence 
of the acoustic power emission on the turbulent Mach number. A Mach-number 
dependence of $\Gamma=5$ assumes that acoustic emission is dominated by the
energy-bearing eddies ; if acoustic emission is predominantly emitted by
inertial-range eddies $\Gamma$ has the value 7.5.
Table~\ref{tab:parameters} lists the values of $\Lambda$ and $\Gamma$ 
that are assumed in the models A1 and A2. The values for $\Lambda$ provide
for a solar model a similar value for the acoustic flux $F_{\rm ac}$ as the
estimates of \citet{Stein68} and \citet{Musielak94}.
For all the models, except for model~B2, we assume for the mixing-length 
parameter the calibrated solar value $\alpha=2.037$; for model~B2 the 
value $\alpha=1.5$ is assumed.
}

Fig.~\ref{fig:HR} displays the locations of these models in the 
colour-magnitude diagram.
Evolutionary tracks (dashed curves) are shown for models with various masses
and are obtained with the CESAM code by \citet{Morel97} as described 
by \citet{Samadi00II}. The transformation from luminosity, effective 
temperature and surface gravity to absolute magnitude $M_{\rm v}$ and 
dereddened colour indices $(b-y)_{\rm o}$ are obtained from the Basel Stellar 
Library \citep{Lejeune98}.
The blue and red edges of the fundamental radial modes (solid curves)
are calculated in the manner of \citet{Houdek99}.
\fig{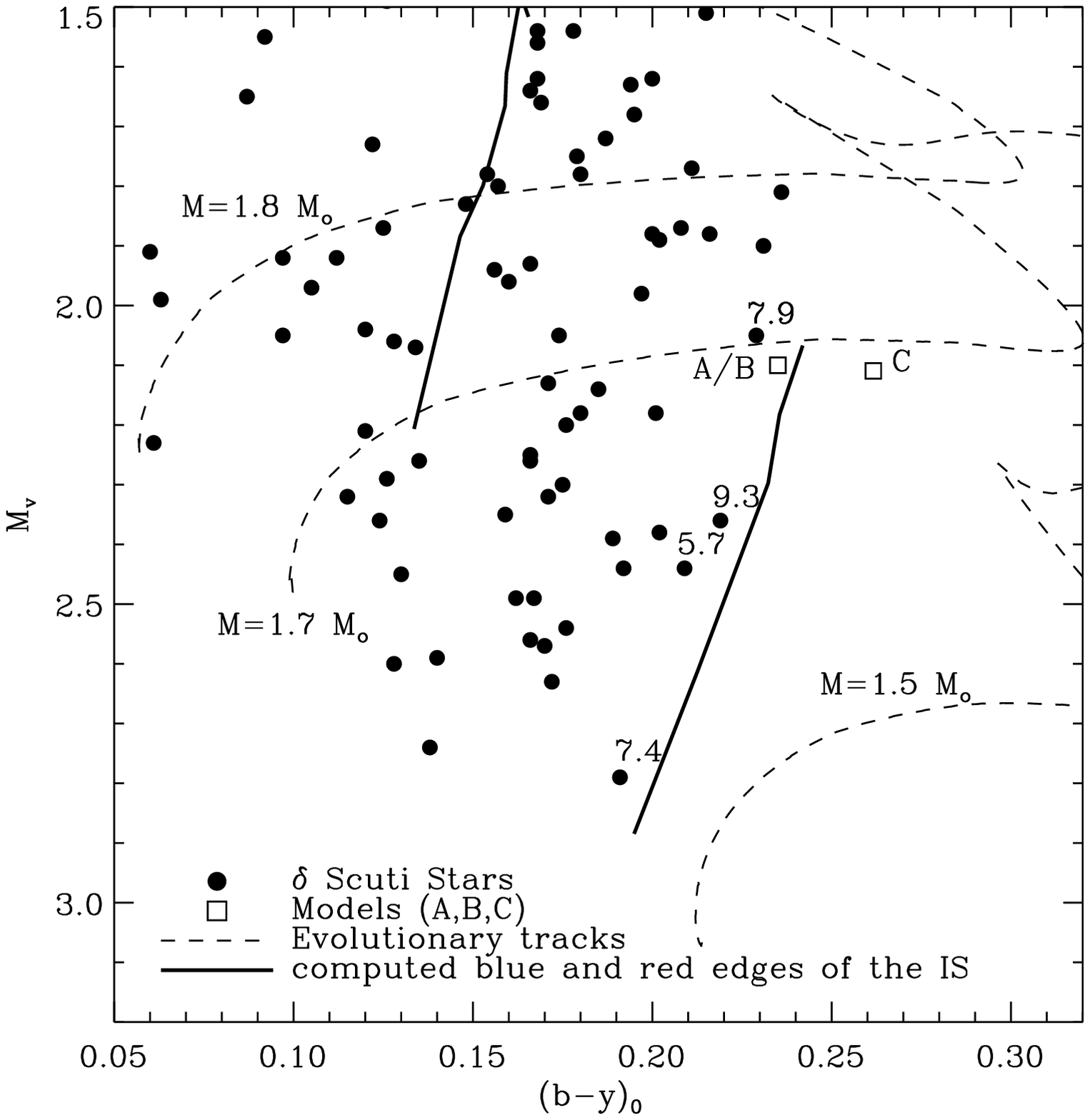}{Colour-magnitude diagram: 
filled circles display the positions of observed $\delta$~Scuti stars from
the \citet{Rodriguez00} catalogue. Squares indicate the positions of the
models A1, A2, B1 ,B2 and C (see Tab.\,1). Dashed curves show evolutionary 
tracks for models with masses 1.5\,M$_\odot$,
1.7\,M$_\odot$ and 1.8\,M$_\odot$. Solid curves display theoretical locations 
of the blue and red edges for the fundamental radial modes according to
\citet{Houdek99}.  Numbers associated with the
symbols indicate apparent magnitudes $V$ for
selected observed $\delta$~Scuti stars. 
}{fig:HR}
The positions of the observed $\delta$~Scuti stars (filled circles) 
are taken from \citet{Rodriguez00}: 
absolute magnitudes, derived from Hipparcos distances and 
dereddened colour indices were kindly supplied by E. Rodr\'iguez (2001,
personal communication; see \citealp{Rodriguez01} , for details).

\section{Stability analysis}

\fig{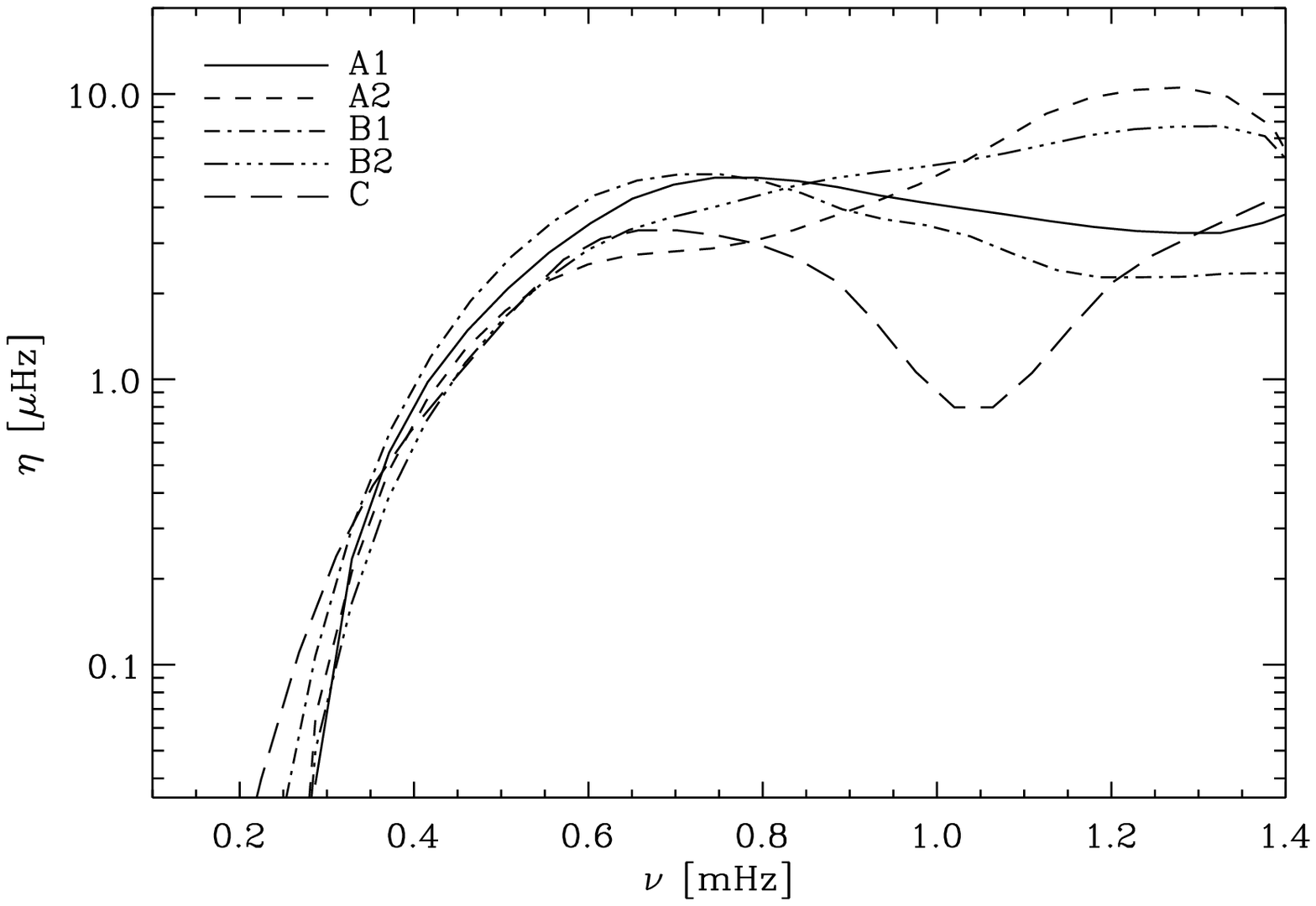}{Damping rates of stable radial p modes as function of 
                       frequency for all stellar models.}
		       {fig:cmp_eta}

The stability computations are as described in 
\citet[$\,\,$and references therein]{Houdek00}. In particular
they include the Lagrangian perturbations of the
turbulent fluxes (heat and momentum) according to Gough's (1977) 
time-dependent formulation. Assuming a temporal dependence, 
$\exp(-{\rm i}\omega t)$, for the pulsations, the complex eigenfrequencies 
of the modes can be written as $\omega=\omega_{\rm r}+{\rm i}\omega_{\rm i}$, 
which defines the cyclic pulsation frequency $\nu=\omega_{\rm r}/2\pi$ and 
the damping/growth rate $\eta=-\omega_{\rm i}/2\pi$.
The outer boundary conditions are applied at the temperature minimum, the
mechanical boundary condition being consistent with a perfectly reflecting 
surface; at the base of the envelope, conditions of adiabaticity and vanishing 
displacement are imposed.

For model~C all the modes are found to be linearly stable (i.e., $\eta>0$) 
as is expected for models lying well outside the IS. This is also found
for the hotter model~B1.
{For the model~A1 (resp. A2) 
the first  four (resp. three) radial
modes, $n$=1,...,4 (resp. $n$=1,2,3), are found to be overstable.}
With the inclusion of a model for the acoustic radiation in the equilibrium 
structure the efficacy with which convection transports the turbulent fluxes 
is decreased \citep[see][]{Houdek98}. This leads to a decrease in the turbulent
Mach number and to a consequent reduction of the stabilizing influence of the 
perturbed momentum flux on the mode damping. The driving 
eventually dominates over the damping leading to overstable modes.

{Reducing $\alpha$ has a similar effect on mode stability than
the inclusion of acoustic radiation in the equilibrium model
\citep[see ][]{Houdek98,Michel99,Houdek00}.
The model B2 was computed with the smaller mixing-length parameter 
$\alpha=1.5$, leading to overstable modes with radial orders $n=$1,2,3.}
 
Table~\ref{tab:unstable} displays the frequency $\nu$ and damping/growth
rate $\eta$ for all overstable radial modes ($\eta<0$) found in the 
models~A1, A2 and B2.

Fig.~\ref{fig:cmp_eta} displays the damping rates as function of frequency
for all stable modes and for all stellar models. 
The coolest model C exhibits a sharp dip in $\eta$ at about 1.1\,mHz,
which we moderated by applying a median filter with a width in
frequency corresponding to nine radial modes; the result is plotted by the
long-dashed curve in Fig.~\ref{fig:cmp_eta}. This pronounced depression in
$\eta$ in model~C is related to the structure of the outer superadiabatic 
boundary layer:
with decreasing surface temperature the location of the superadiabatic 
boundary layer is shifted progressively deeper into the star. This modifies
the thermodynamic properties of this boundary layer of finite thickness, in 
particular, the thermal relaxation time \citep{Balmforth92a}. The thermodynamic 
coupling between the pulsations and the superadiabatic boundary layer becomes 
more efficient in cooler models, thereby promoting the depression in the 
damping rates by radiative processes \citep[see][]{Houdek99}.

\begin{table}
\caption{
Frequency $\nu$, damping/growth rate $\eta$ and stability coefficient 
$\omega_{\rm i}/\omega_{\rm r}$ for all overstable radial p modes
predicted for the models~A1, A2 and B2.}
\label{tab:unstable}
\begin{center}
\begin{tabular}{cccccc}  
Model&n & $\nu$    &$\eta$&$\omega_{\rm i}/\omega_{\rm r}$\cr
&  & [$\mu$Hz]&[nHz] &$\times10^{-6}$                \cr
\hline
   & 1 & 123 &-0.03 &0.25\cr
A1 & 2 & 161 &-0.31 &1.92\cr
   & 3 & 202 &-4.14 &20.48\cr
   & 4 & 244 &-3.90 & 15.97 \cr
\hline
   & 1  & 124 & -0.04 & 0.36       \cr
A2 & 2  & 162 & -0.40 & 2.47       \cr
   & 3  & 203 & -1.27 &  6.25 \cr
\hline
   & 1  & 124 & -0.04 &  0.34 \cr  
B2 & 2  & 161 & -0.31 &  1.95 \cr  
   & 3  & 203 & -0.83 &  4.09 
\end{tabular}
\end{center}
\end{table}

\section{Excitation rate and amplitude spectrum}
The rms value of the mode surface velocity, $v_{\rm s}$, is related to the 
damping rate, $\eta$, and to the rate at which energy is injected into the 
mode (excitation rate), $P$, by
\eqn{
v_{\rm s}^2 = \xi_{\rm r}^2(r_{\rm s})\;\frac{P}{2\,\eta\,I}\,,
\label{eqn:vs2}
}
where 
$\displaystyle{\xi_{\rm r}}$ is the radial displacement eigenfunction, 
$r_{\rm s}$ is the radius at which the surface velocities are measured
and which we assume to be 200\,km above $T_{\rm eff}$,
and the mode inertia $I$ satisfies
\eqn{
I=\int_0^M\vec\xi^*\cdot\vec\xi\,{\rm d}m\,.
\label{eqn:inertia}
}
The rate of energy injected into a mode is computed according to 
Paper~I and is proportional to
\eqna{
P(\omega) & \propto & 
\int_{0}^{M}\rho\,w^3\,\ell^4\,\left
(\frac{{\rm d}\xi_{\rm r}}{{\rm d}r}\right)^2\mathcal{S}(\omega,m)\,{\rm d}m\,,
\label{eqn:A2}
} %
where $\rho$ is the density, $\ell$ is the mixing length, and $w$ is the 
vertical component of the rms velocity of the convective elements.  
The function $\mathcal{S}(\omega,m)$ describes approximately contributions from
eddies with different sizes to the excitation rate $P$.
Detailed expressions for $\mathcal{S}(\omega,m)$ were given in Paper~I.
For estimating ${\cal S}$, assumptions for the turbulent kinetic energy 
spectrum $E(k)$, and for the turbulent spectrum of the entropy fluctuations 
$E_{\rm s}(k)$ have to be made, where $k$ is the eddy wavenumber.
In this paper we assume for $E(k)$  the ``Nesis Kolmogorov Spectrum'' 
(NKS hereafter) as discussed in Paper~I. This turbulent spectrum is obtained
from observations of the solar granulation by \citet{Nesis93}, and leads to 
the best agreement between a solar model using our stochastic excitation theory 
and solar measurements \citep{Samadi01b}. 
\fig{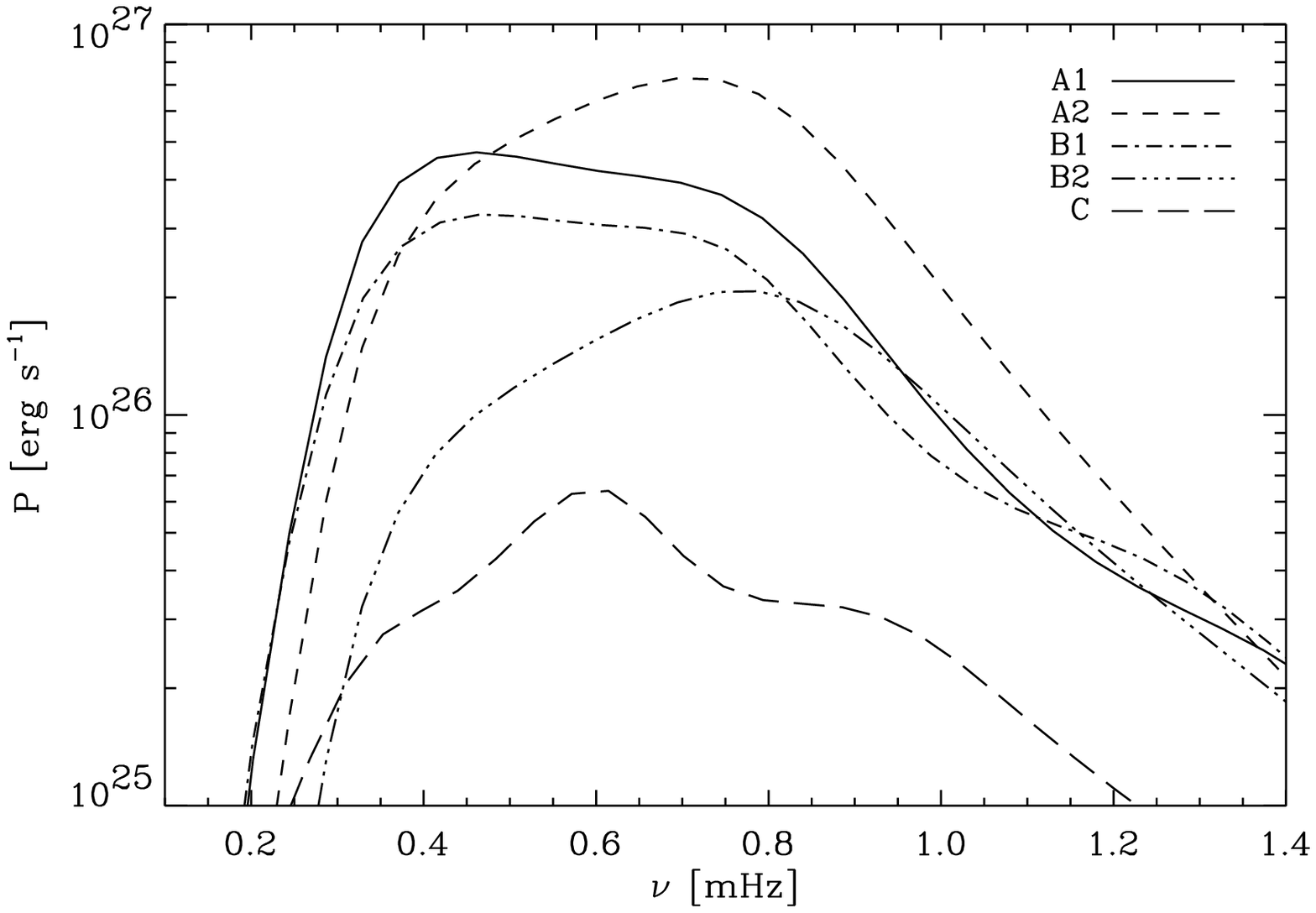}
    {Excitation rate $P$ as function of frequency for all stellar models.}
{fig:cmp_pow}
   
\fig{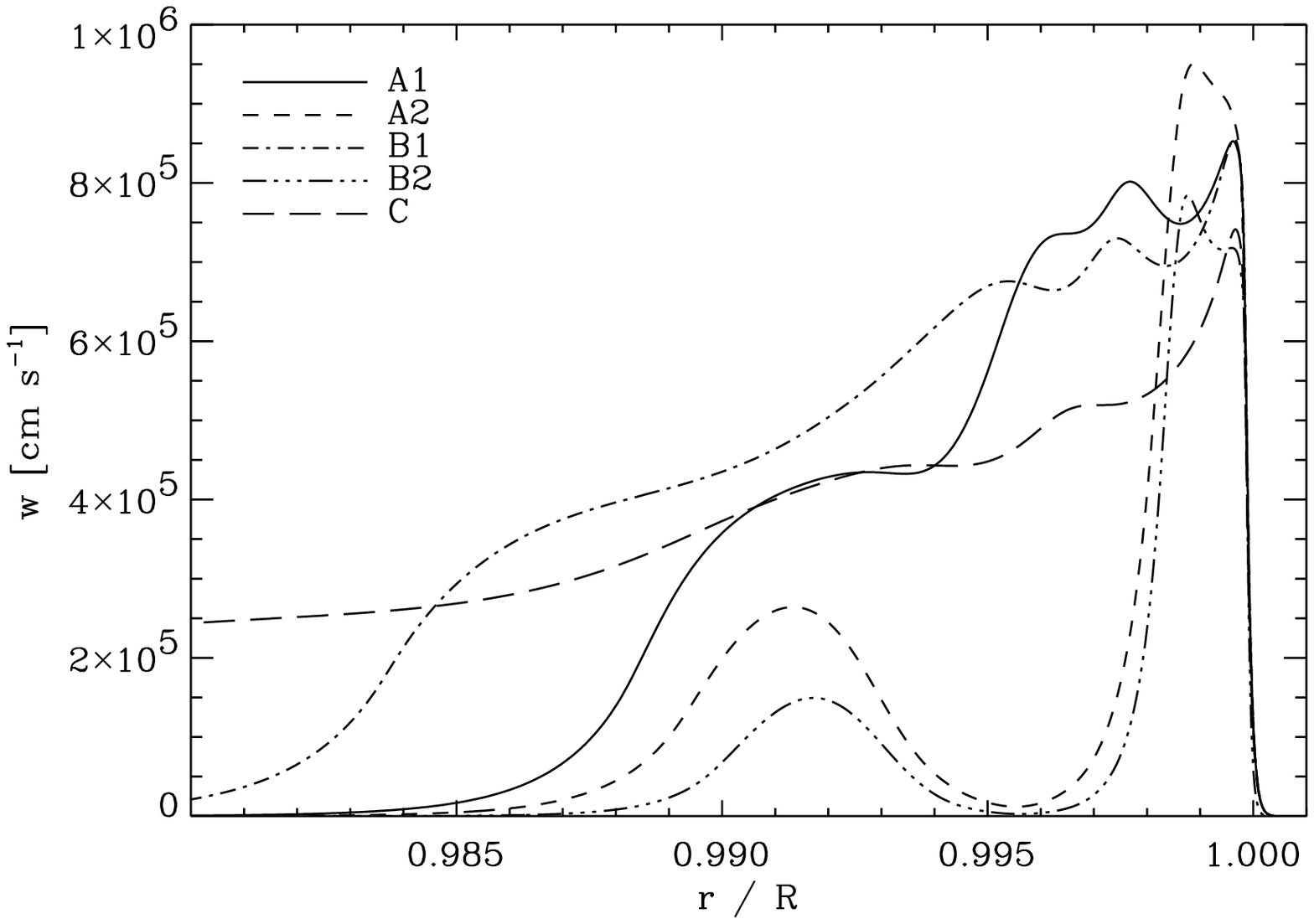}
    {Depth dependence of the vertical component of the convective velocity,
     with $R$ being the radius at the photosphere ($T=T_{\rm eff}$).
}
{fig:cmp_profil_w}

Results for the estimated excitation rate $P$ are depicted in 
Fig.~\ref{fig:cmp_pow}. For the models~A1, A2 and  B1 the excitation rate 
$P$ is about one magnitude larger than for model~C. This is a result of 
the larger convective velocities in the superadiabatic boundary layers of  
the models~A1, A2 and B1 (see Fig.~\ref{fig:cmp_profil_w}), which are all
hotter than model~C.
{The frequency dependence of $P$ for the models A1 and B1 are similar as it is
also the case for the models A2 and B2; the latter two models, however, are 
quite different from the first two models.
This difference in the frequency dependence of $P$ is a consequence of 
the different profiles of the convective velocities $w$ 
(see Fig.~\ref{fig:cmp_profil_w}); the profiles of $w$ are similar for 
A2 and B2 but differ substantially from A1 and B1. For the models
A2 and B2 the efficacy of convection has been reduced severely by either 
including acoustic radiation in the equilibrium model (A2) or by reducing
the mixing-length parameter $\alpha$ to a value much smaller than the
calibrated value for a solar model (B2). This
results in shallower superadiabatic regions and in 
larger superadiabatic temperature gradients; pulsation modes in A2 and B2
are therefore predominantly excited at the very top of the convection zone, 
whereas in the models A1 and B1 the modes are excited over a larger driving 
region. 
The two sets of values for $\Lambda$ and $\Gamma$, 
listed in Table~2, provide in a solar model approximately the 
same value for the acoustic flux $F_{\rm ac}$ (see Section~2).
In model A2 (which assumes $\Lambda=2000$ and $\Gamma=7.5$), however, 
$F_{\rm ac}$ is about three times larger than in model 
A1 (which assumes $\Lambda=100$ and $\Gamma=5$), and the associated 
velocities, plotted in Fig.~\ref{fig:cmp_profil_w}, are correspondingly 
greater.
}

%

{It is interesting to note that in Fig.~\ref{fig:cmp_profil_w} the 
convective velocities in the superadiabatic boundary layers of models A1 
and A2 are somewhat larger than the others.  Although including acoustic 
radiation in the mean stellar structure leads to a reduction of the maximum 
turbulent Mach number $M_{\rm t}\equiv w/c$ (by $\sim 1.5\%$ for model A1 and
$\sim 5\%$ for model A2 relative to B1), the whole structure of the outer 
layers changes too, thereby increasing the locally defined adiabatic sound 
speed $c$. This leads to an increase of the convective velocities $w$ in the
very outer layers despite the decrease in $M_{\rm t}$.}

The less smooth frequency dependence (wiggles) of $P$ for $\nu\gtrsim0.7~\mu$Hz
for model~B1 (and model~C) is related to the location and extent of the 
driving region: 
the radial eigenfunctions $\xi_{\rm r}$ vary rapidly with depth and frequency 
(particularly for the high-frequency modes). As discussed above, in the models 
A2 and B2 the modes are predominantly excited in a shallow region beneath the 
surface, where the expression $({\rm d}\xi_{\rm r}/{\rm d}r)^2$ 
(see Eq. \ref{eqn:A2}) varies monotonically with frequency 
\citep{GMK94,Samadi00III}, leading to the smooth frequency dependence of $P$ 
for $\nu \gtrsim 0.7~\mu$Hz, as depicted in
Fig.\,\ref{fig:cmp_pow}. The larger driving regions in the models B1 and C
extend to layers where the expression $({\rm d}\xi_{\rm r}/{\rm d}r)^2$ no 
longer varies monotonically with frequency leading to the frequency-dependence 
of $P$ as shown by the dot-dashed and long-dashed curves in 
Fig.\,\ref{fig:cmp_pow}.

\begin{figure}[tbp]
\vspace{-2mm}
\resizebox{\lenA}{!}{\includegraphics  {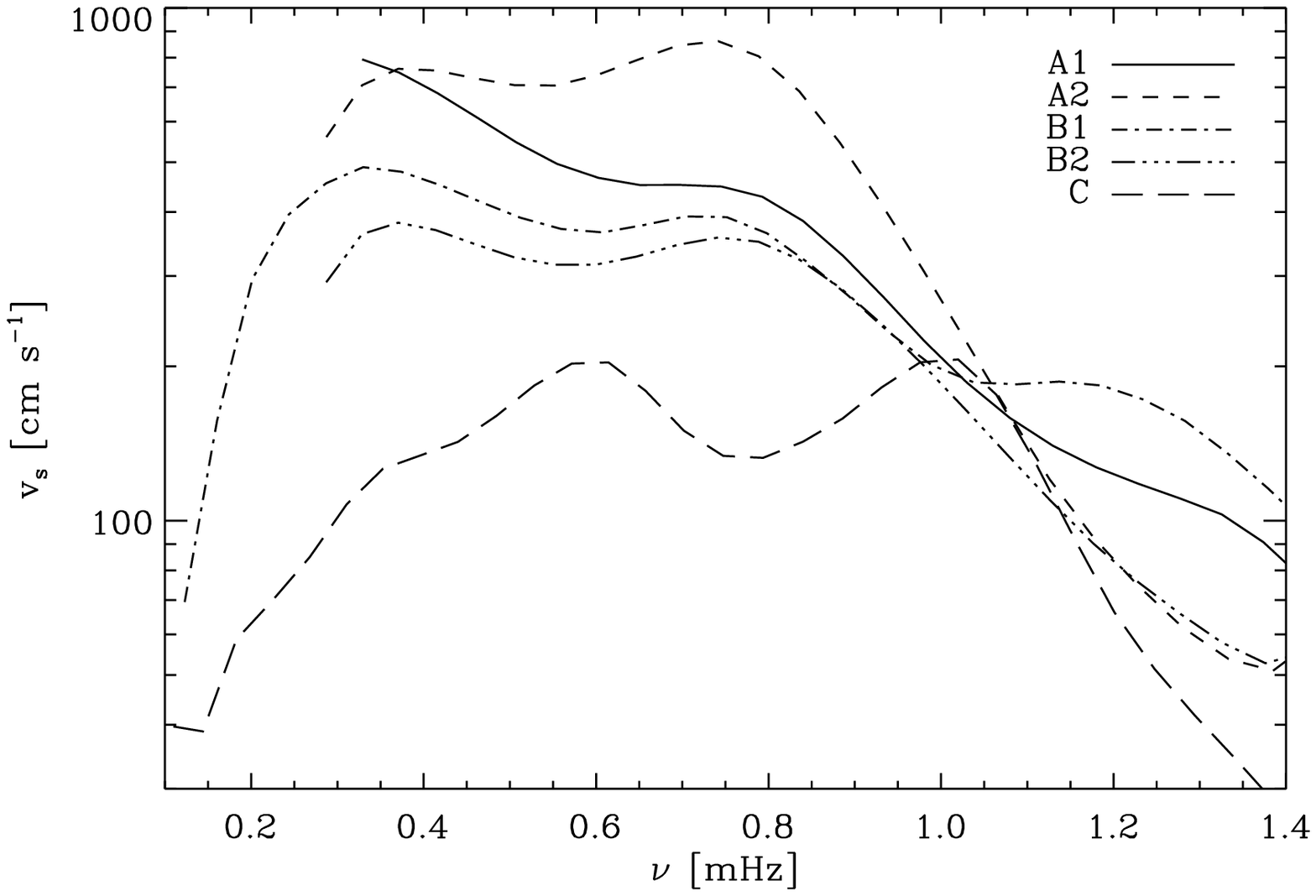}}
\vspace{-2mm}
\resizebox{\lenA}{!}{\includegraphics  {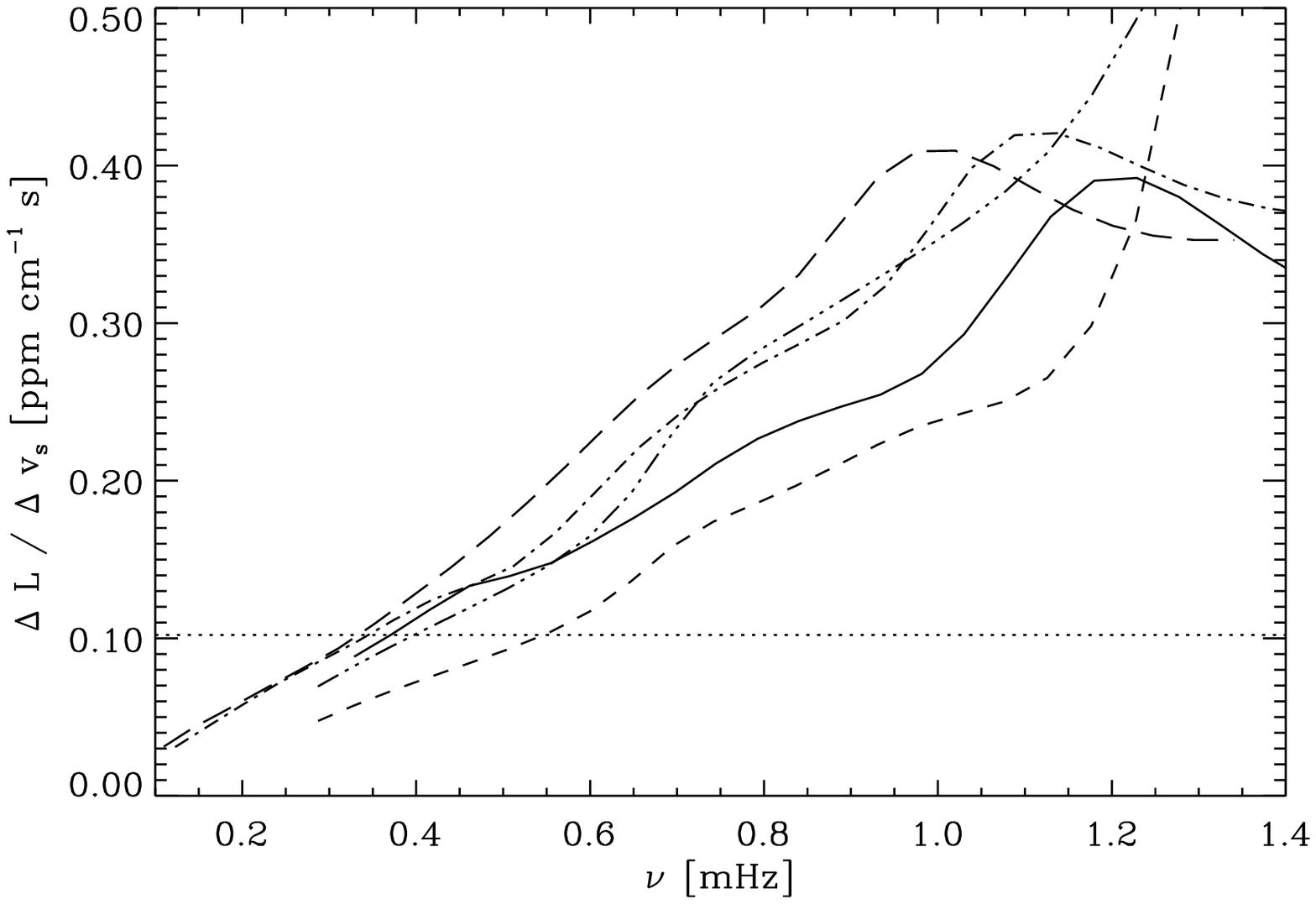}}
\vspace{-1mm}
\resizebox{\lenA}{!}{\includegraphics  {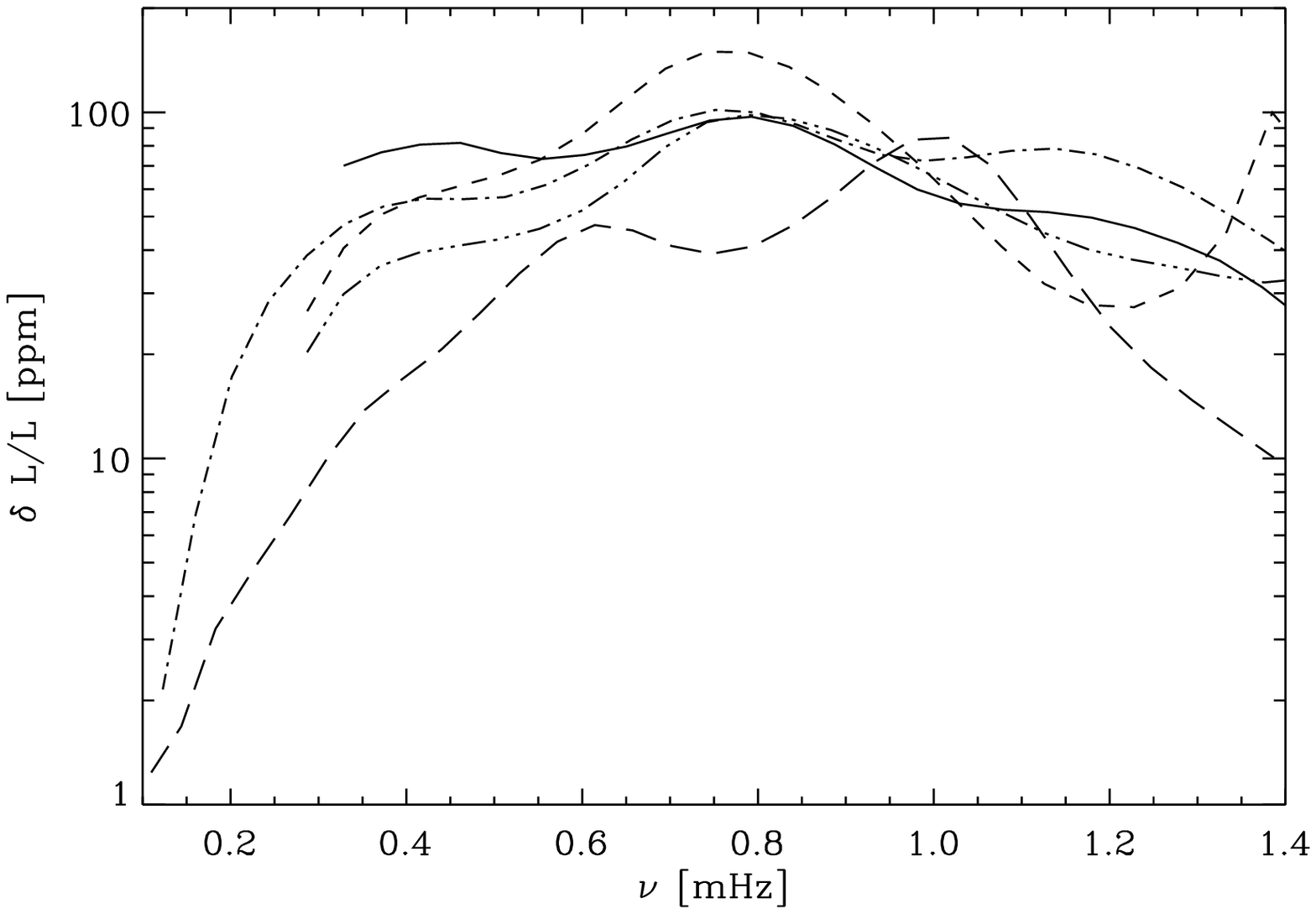}}
\vspace{-4mm}
\caption{Linear oscillation amplitudes of stable radial modes as a function of
         frequency. The
         {\bf top panel} displays the surface velocity amplitudes and the
         {\bf middle panel} the amplitude ratios, luminosity over velocity 
	 amplitudes, computed 200\,km above the photosphere ($T=T_{\rm eff}$). 
	 The dotted horizontal line represents the results for 
	 Kjeldsen \& Bedding's (1995) scaling expression (\ref{eqn:dl_vs}), 
	 assuming $T_{\rm eff}=6839\,$K.
	 In the {\bf bottom panel} the luminosity amplitudes are depicted.
	 In the middle and bottom panel the luminosity amplitudes are
	 computed at the outermost meshpoint of the models.
         \vspace{-10mm}
	 }
\vspace{0.25cm}
\label{fig:cmp_ampli}
\end{figure}

In the top panel of Fig.~\ref{fig:cmp_ampli} the surface velocity amplitudes 
$v_{\rm s}$ are depicted for all stellar models, computed according to 
Eq.(\ref{eqn:vs2}). 
{In the models~A1, A2, B1 and B2 the amplitudes of stochastically excited 
p modes are larger ($\sim 5-9~{\rm m s}^{-1}$)
than in model~C ($\sim 2~{\rm m s}^{-1}$).}

For estimating the luminosity amplitudes the full nonadiabatic
luminosity eigenfunctions have to be used. The relative luminosity amplitudes,
$\delta L/L$, are linearly related to the velocity amplitudes, i.e. they
are proportional to the ratio of the luminosity eigenfunction over the
displacement eigenfunction. This ratio is determined by the solution of
the nonadiabatic pulsation equations and is independent of a stochastic
excitation model \citep[see][]{Houdek99}. In the middle panel of 
Fig.~\ref{fig:cmp_ampli} 
the amplitude ratios, $\Delta L/\Delta v_{\rm s}$, are plotted as a function of 
frequency for all stellar models. The shape of the amplitude ratios are in 
general 
similar between all the models with the smallest ratios predicted for the 
models~A2 and A1. Only at the highest frequencies the amplitude ratios are
considerably larger in A2 and B2; at high frequencies nonadiabatic effects
due to radiative dissipation in the radiative zone, below the shallow surface
convection zones in A2 and B2, lead to an increase in the amplitude of the 
luminosity eigenfunctions and consequently in the luminosity amplitudes. 
The velocity amplitudes in Fig.~\ref{fig:cmp_ampli} are 
obtained 200\,km above the photosphere ($T=T_{\rm eff}$) and do increase
by a factor of about two at the outermost meshpoint of the model, i.e. at
an optical depth $\tau=10^{-4}$.

{We predict a maximum value of the luminosity amplitude 
$\delta L/L\sim 97$~ppm for model~A1, 
$\delta L/L\sim 150$~ppm for model~A2, $\delta L/L\sim 101$~ppm for model~B1 , 
$\delta L/L\sim 98$~ppm for model~B2  and $\delta L/L\sim 84$~ppm for 
model~C.  These results are summarized in Table~\ref{tab:ampli}. }

The dotted horizontal line in the middle panel of Fig.\,\ref{fig:cmp_ampli}
represents an order-of-magnitude estimate of the amplitude ratio according 
to \citet{Kjeldsen95}:
\eqn{		
\delta L/ L  \propto v_{\rm s} \,  T_{\mathrm{eff}}^{-1/2}\,.
\label{eqn:dl_vs}
}
\citeauthor{Kjeldsen95} derived this expression for a purely radiative model
assuming simplified proportional relations in the adiabatic approximation.
{ 
This simplified scaling law suggests smaller values for the amplitude 
ratios and consequently leads to smaller luminosity amplitudes $\delta L/L$, 
particularly at high frequencies, where nonadiabatic effects are important. At
a frequency $\nu \simeq 1$~mHz, for example, the scaling law~(\ref{eqn:dl_vs})
predicts for model A1 a luminosity amplitude which is about three times smaller than that obtained from the nonadiabatic computation.
} 

There is evidence that energy equipartition holds for the Sun (apparently
fortuitously); an estimate of the total energy in the modes is, however, only
possible for the Sun for which accurate data are available; using GONG data
the total energy for modes with degrees $l=0,...,300$ and with radial
order $n$ up to 10 is found to be approximately $2\times10^6~E_0$, where
$E_0 \sim 2\times10^{28}~$ erg is the maximum value of the kinetic energy in a
particular ridge (i.e. for a particular value of $n$) and which is independent
of $l$ \citep[see Fig.\,\,5 of\,][]{Komm00}.
The value $2\times10^6$ is also roughly equal to the number of granules on 
the solar surface, a result which supports the energy equipartition principle.
In other stars, however, energy equipartition does not necessarily hold,
because we have a nonequilibrium dynamical (yet statistically steady) system
in which the damping and excitation is balanced in a nonlinear way by the
energy input and output, i.e. it is not determined by equilibrium. In such
a nonequilibrium situation there is no general physical principle limiting
the ratio of the energy in the oscillation mode to the energy in the
convection. Another facet of such a dynamical process is provided by the
reaction of convection to the acoustical radiation; although the latter
contributes towards augmenting the damping of an eddy, the resultant change
of the background stratification is such as to augment the driving by even
more, causing the convective velocities to increase.

{
It is perhaps interesting to mention that the acoustic energy flux 
generated by the fluctuating Reynolds stress of the turbulent velocity field
is relatively small compared to the total energy flux carried by the 
convection; the ratio between the acoustic energy flux emitted by the 
energy-bearing eddies and the convective energy flux 
is proportional to $M_{\rm t}^{\Gamma}$ with $\Gamma=5$ for homogeneous, 
isotropic turbulence \citep[see][]{Lighthill52}. The turbulent Mach number 
in $\delta$ Scuti stars is in general much smaller than unity and consequently 
this ratio is small.  For the Sun this ratio is of the order of 
$\sim10^{-3}$ \citep[see e.g., ][] {Stein68}.
The acoustic flux emitted predominantly by inertial-range eddies is 
proportional to $M_{\rm t}^{\Gamma}$ with $\Gamma=15/2$
\citep[see][]{Goldreich90}, {\it i.e.} it scales with an even higher power
of the Mach number (see also Section~2). Consequently the total amount of 
acoustic energy injected into the p modes is small compared to the energy 
carried by the convection.
In a fully convective envelope the total energy flux (luminosity) is carried
solely by the turbulent velocity field, {\it i.e.} in that case the luminosity 
is a measure of the total energy in the convection. Therefore the ratio between
the energy supply rate for a particular mode and the luminosity, $P/L$, is 
proportional to the ratio between the energy in that mode and the total energy
in the convection. In the Sun this ratio is of the order of $\sim 10^{-11}$ for
the mode with the largest amplitude. In model A2 this ratio is $\sim 10^{-8}$,
which is still small.
}

\begin{table}
\caption{{Maximum values of the estimated velocity, $v_{\rm s}$, and luminosity, 
$\delta L/L$, amplitudes.}}
\label{tab:ampli}
\begin{center}
\begin{tabular}{ccc}  
Model & $v_{\rm s}$  & $\delta L/L$ \cr
      & $[{\rm~m s}^{-1}]$  & [ppm]\cr
\hline
A2  &     8.6  &  150 \cr
A1  &     7.9     & 97  \cr
B1  &    4.9     & 101\cr
B2  &    5.5     & 98\cr
C   &    2.0     & 84\cr
\end{tabular}
\end{center}
\end{table}

\section{Observational constraints for detecting solar-type oscillations}

There have been recent reports on the possible detection of
solar-type oscillations in $\alpha$~Cen (HD~128620) by
\citet{Bouchy01}, in $\beta$~Hydri (HD~2151) by \citet{Bedding01} and in 
Procyon~A (HD~61421) by \citet[][ see also \citealp{Barban99}]{Martic99},
who obtained spectroscopic surface velocity measurements of these bright stars 
(the apparent magnitude $V=2.80$ for $\beta$~Hydri, $V=0.34$ for Procyon and 
$V=-0.1$ for $\alpha$~Cen) from the ground.
The maximum values of the observed peak-velocity amplitudes are of the order 
$\sim 35~{\rm cm s}^{-1}$ for $\alpha$~Cen,  $\sim 50~{\rm cm s}^{-1}$ 
for $\beta$~Hydri and $\sim 50~{\rm cm s}^{-1}$ for Procyon.
Current ground-based instruments are able to detect oscillations with velocity
amplitudes of the order predicted for our models~A1, A2 and~B1, B2, but 
only for stars with an apparent  magnitude $V$ of less than 
$\sim 3-4$ (Bouchy, personal communication). 
The HARPS (High-Accuracy Radial-velocity Planetary Search) project 
\citep{Bouchy01}, for example, will be able to detect oscillations 
with our predicted velocity amplitudes for stars with an apparent magnitude 
smaller than $\sim 4-5$. This detection threshold is still too small for 
detecting solar-type oscillations in $\delta$~Scuti stars 
located near the red edge of the IS, particularly in view of the fact that 
most of the currently known $\delta$~Scuti stars are even fainter. 
For example, the apparent  magnitudes of known 
$\delta$~Scuti stars located nearest to the red edge (see Fig.~\ref{fig:HR}) 
are between $V=5.7$ and $V=9.3$. 

Future space missions with instruments dedicated to asteroseismology, however,
will be able to detect solar-like oscillations in $\delta$~Scuti stars:
the forthcoming space project COROT \citep{Baglin98}, for example,
will reach a noise level of $0.7$ ppm \citep{Auvergne00} for a star with an
apparent magnitude of $V=6$, using photometric measurements.
Therefore, in stars with similar magnitudes, COROT will be able to detect 
oscillation amplitudes as small as $\sim 3\,$~ppm, a value which is 
similar to that measured in the Sun. 
The instrument on COROT will be limited by the photon noise only for stars 
with magnitudes larger than $V \simeq 9$: i.e., for a star with magnitude 
$V\simeq 8$ the detection threshold will be $\sim 5$~ppm. 
This threshold is small enough to detect and measure 
many solar-like oscillations in $\delta$~Scuti stars which are similar 
to the $\delta$~Scuti models considered in this paper.

 
\section{Conclusion}

We studied oscillation properties in $\delta$~Scuti stars located near
the observed red edge of the classical instability strip.
Such stars can pulsate with both opacity-driven modes and intrinsically stable 
stochastically driven (solar-like) modes.
The estimated velocity amplitudes of the stochastically driven modes in our 
$\delta$~Scuti models are found to be larger than in cooler and pulsationally
stable models lying outside the IS. This result supports the idea that
solar-like oscillations in $\delta$~Scuti stars may be detected.

Including a model for the acoustic radiation in the equilibrium model results
in a cooler red edge and does effect the properties of the excitation rate
of p modes (see also \citealp{Houdek98}, \citealp{Houdek00}); in particular
the pulsation amplitudes do become larger and are predicted to be largest
for a model with the largest acoustic flux $F_{\rm ac}$ (i.e., model A2).
{Moreover, for the $\delta$~Scuti models considered in this paper, 
overstable modes were predicted only if either acoustic emission in the mean 
stratification was included or if the mixing-length parameter was reduced to 
a value smaller than suggested by a calibrated solar model.}

A potential target star should neither be too cool
(i.e., no opacity-driven modes) nor too hot (i.e., stochastically excited 
modes with amplitudes too small to be detectable). 
We quantify this with the illustrative case of our 
$\delta$~Scuti models with a mass $M =1.68\,$M$_\odot$ and
we identify the following $\delta$~Scuti stars from the \citet{Rodriguez00}
catalogue, located near the red edge, as potential candidates
for the target selection of upcoming observing campaigns:
HD57167, HD14147, HD208999 and HD105513.  

Although the amplitudes of the solar-type oscillations, predicted in
our $\delta$~Scuti models, are large enough to be detected from ground, 
today's ground-based instruments will detect such oscillations only in 
brighter $\delta$~Scuti stars with an apparent magnitude of 
up to $V\sim 3-4$ (Bouchy, 2001, personal communication).
However, new ground-based observing campaigns, such as 
the HARPS project \citep{Bouchy01} will be able to detect stochastically
excited oscillations in  $\delta$~Scuti  stars with an apparent 
magnitude of up to $V \sim 4-5$. 
Unfortunately, there are no such bright stars in the \citet{Rodriguez00} 
catalogue which are located near the red edge, although some bright stars 
near the red edge may have opacity-driven modes with amplitudes too small 
to be detectable with today's ground-based instruments and are therefore 
not classified as $\delta$~Scuti stars.

The forthcoming space missions for asteroseismology, such as COROT 
\citep{Baglin98} and Eddington \citep{Favata00} will be able to detect 
solar-like oscillations in faint $\delta$~Scuti stars. 
The large instrument on the Eddington spacecraft will measure 
stellar oscillations with amplitudes as small as $1.5$~ppm in stars with an 
apparent magnitude of $V \simeq 11$ assuming an 
observing period of 30 days.
Moreover, Eddington's large field of view 
will allow it to monitor a large number of stars simultaneously.
This will be helpful for detecting and classifying new $\delta$~Scuti
stars and for measuring the location of the red edge of the IS with greater
precision than it was possible before.

\begin{acknowledgements}
We thank E. Rodr\'iguez for providing the $\delta$~Scuti data set in a
convenient and immediate usable form, T. Lejeune for allowing us to use the 
Basel library and D. Cordier for providing it on the Internet.
We thank A. Baglin for useful discussions on the COROT specifications,
F. Bouchy for providing valuable information on the HARPS project and related 
experiments, and C. Catala and E. Michel for useful discussions on the
possibilities of detecting new $\delta$~Scuti stars. 
We are grateful to Douglas Gough for very helpful discussions on stochastic
mode excitation and to Mike Montgomery for improving the English.
GH and RS acknowledge support by the Particle Physics and Astronomy 
Research Council of the UK.
RS's work has been supported under the grant PPA/G/O/1998/00576.
\end{acknowledgements} 



\bibliographystyle{aa}
\end{document}